\UseRawInputEncoding
\documentclass[a4paper,11pt]{article}

\pdfoutput=1 % if your are submitting a pdflatex (i.e. if you have
\usepackage{jcappub} % for details on the use of the package, please
\usepackage[T1]{fontenc} % if needed

\title{\boldmath 
StellarICS: Inverse Compton Emission from the Quiet Sun and Stars from keV to TeV}

\author[a,b]{Elena Orlando\note{orlandele@gmail.com}}
\author[c]{and Andrew Strong}
\affiliation[a]{University of Trieste and Istituto Nazionale di Fisica Nucleare, Italy}
\affiliation[b]{Kavli Institute for Particle Astrophysics and Cosmology and Hansen Experimental Physics Laboratory, Stanford University, CA, USA}
\affiliation[c]{Max-Planck-Institut f\"ur extraterrestrische Physik, Germany}
\abstract{The study of the quiet Sun in gamma rays started over a
decade ago, and rapidly gained a wide interest. 
Gamma rays from the quiet Sun are produced by Galactic Cosmic Rays (CRs) interacting with its surface (disk component) and with its photon field (spatially extended inverse-Compton component, IC). The latter component is maximum close to the Sun and it is above the background even at large angular distances, extending over the whole sky. First detected with EGRET, it is studied now with Fermi-LAT with high statistical significance. 

Observations of the IC component allow us to obtain information on CR electrons and positrons close to the Sun and in the heliosphere for the various periods of solar activity and polarity. They allow to learn about CR interactions and propagation in the stellar photosphere and heliosphere, and to understand the solar environment and its activity.  Analyses of solar observations are usually model-driven.
Hence advances in model calculations and constraints from precise CR measurements are timely and needed. 

Here we present our StellarICS code to compute the gamma-ray IC emission from the Sun and also from single stars. 
The code is publicly available and it is extensively used by the scientific community to analyze Fermi-LAT data. It has been used by the Fermi-LAT collaboration to produce the solar models released with the FSSC Fermi Tools. 
Our modeling provides the basis for analyzing and interpreting high-energy data of the Sun and of stars. 

After presenting examples of updated solar IC models in the Fermi-LAT energy range that account for the various CR measurements, we extend the models to keV, MeV, and TeV energies for predictions for present and future possible telescopes such as AMEGO, GECCO, an e-ASTROGAM-like instrument, HAWC, LHAASO, SWGO, and X-ray telescopes. We also present predictions for some of the closest and most luminous stars.\\      
}

\begin{document}
\maketitle
\flushbottom

\section{Introduction}
\label{sec:intro}

The Sun was observed as steady state gamma-ray source over a decade ago. Specifically, by using archival data of the EGRET  \cite{hartmann} detector on NASA's Compton Gamma Ray Observatory GCRO in 2008, we  \cite{Orlando2008} obtained the first detection of the gamma-ray emission from the Sun in its non-flaring state. In that work two different spatial and spectral components of the quiet solar emission were distinguished: the hadronic component from the disk, and the leptonic component with an angular extension around  the solar disk. 
Because of its association with Galactic cosmic rays (CRs), the gamma-ray brightness of both components was predicted to vary over the solar cycle and was predicted to anti-correlate with the solar activity. 

Before its detection \cite{Orlando2008}, the disk emission was briefly mentioned by \cite{hudson} in 1989, while 
in 1991 the work in \cite{Seckel91} contains the first calculation of the gamma-ray emission from pion-decay by CR proton cascades in the solar atmosphere, and in 1997 an upper limit with EGRET data was obtained by \cite{Thompson}. Further studies of this component followed also recently both from the theoretical point of view \cite{Mazziotta, Li, Becker, Guti, Hudson20, Nibl, Zhou} and from the observational one (\cite{Abdo2011, Barbiellini, Bartoli, Linden, HAWC, Linden2020, Ng, Tang}). Regarding the solar gamma-ray leptonic component, in 2006 we \cite{Orlando2006} predicted the existence of an extended inverse Compton (IC) emission from scattering by Galactic CR electrons and positrons (all-electrons thereafter) on solar photons. In the same period, an independent group \cite{Moskalenko} made similar predictions. This emission, which is produced in the entire heliosphere, is brighter for directions close to the Sun,  and has a broad distribution, effectively covering the whole sky.  Even at large elongation angles its brightness is comparable to the brightness of the diffuse extragalactic emission. Observations of the IC emission from the Sun and comparison with models can probe the CR all-electron spectra at different distances throughout the inner heliosphere, and allow studies of the solar modulation even at  close proximity to the Sun. 
After the first detection of this component \cite{Orlando2008}, updated observations of the IC emission were published in \cite{Abdo2011, Raino} using Fermi-LAT \cite{Fermi} data, with much higher statistical significance than EGRET.  
Since then no further analyses results on the IC component from the Sun have been reported in the literature, also due to the limitation of the models that are crucial in the analysis. 
For this reason any improvement in the modeling of the IC component is fundamental and timely.

Modeling the IC emission is also important because the Sun is moving across the sky, and this component acts as confusing foreground for any Fermi-LAT studies. Hence it has to be precisely accounted for  (\cite{OrlandoProc, GulliOrlando}) when analyzing the Fermi-LAT data, as reported in the official Fermi guidelines\footnote{https://fermi.gsfc.nasa.gov/ssc/data/analysis/scitools/solar\_template.html}.\\
In our previous work \cite{Orlando2006} the same mechanism of IC gamma-ray emission from the Sun was predicted to be produced also around other stars, and it was found to be important especially from the closest luminous stars and OB stellar associations. Our model of IC emission from the Cygnus OB2 association \cite{OrlandoProc} was included in the analysis in the study of the Cygnus region \cite{cygnus} that showed freshly accelerated CRs. 

In this paper we  present the StellarICS code to compute the IC emission from the Sun and stars. It is available to the community and can be freely downloaded\footnote{https://gitlab.mpcdf.mpg.de/aws/stellarics/}. Some examples of usage and a sample of models of IC emission from the Sun and stars are shown and described. This code has been used within the Fermi-LAT collaboration for generating the models and by several other independent groups (e.g. \cite{Zhou, Linden2020}). It has been also used to produce the model for the Solar Science Tools within the Fermi Science Tools distributed in the Fermi Science Support Center website. This work will be a reference paper for the code and for the models used for recent and future Fermi-LAT analyses, for future gamma-ray data at MeV and also for analyses of water Cherenkov telescopes' data. Calculations at keV, MeV and TeV energies are also presented in view of forthcoming keV/MeV telescopes and observations (such as AMEGO \cite{AMEGO}, GECCO \cite{Gecco}, an e-ASTROGAM-like instrument \cite{eastrogam}, eRosita \cite{eRosita}) and more sensitive observations at TeV energies (i.e. with HAWC \cite{HAWC}, LHAASO \cite{LHAASO}, and SWGO \cite{SWGO}).

\section{StellarICS: Stellar IC scattering code}
\label{stellarics}
This section contains the technical description of the StellarICS code. Galactic CRs that penetrate the solar (and stellar) system produce gamma-ray emission via IC scattering on the photons generated by the Sun (and stars). StellarICS computes the IC scattering in the heliosphere and photosphere of individual stars for a given CR all-electron spectrum and modulation, and for a given photon spectral and spatial distribution. Because the thermal photon spatial and spectral distributions from the Sun are precisely known, the code needs only assumptions about CRs at different distances from the Sun (or from the star) to calculate the model of the IC emission. \\
StellarICS is written in C++ and  is modular. It is composed by the following C++  classes: IC cross sections, emissivity spectra, stellar radiation fields, electron and positron spectra. StellarICS can compute  both isotropic and anisotropic IC scattering. 
A driver routine is provided to specify the parameters, which can be adapted by the user as required. Moreover, new models of electrons, positrons and their modulation can be easily added by the user. It is very flexible because it contains user-defined parameters (e.g. energy ranges, integration steps). 
StellarICS can be run in parallel and has an optimized emissivity computation for spectral integrations.
 The computation parameters allow a user-defined compromise between high resolution  and reasonable computation time.
Output is provided as FITS files (in table extension format), in various forms: angular profiles, spectra and differential and integrated flux (for more details see Section~\ref{sec26}). A sample output dataset is provided with the package to  illustrate the format and to test the installation. Output as $idl$ commands  is also provided. StellarICS has been tested with GNU and Intel compilers, and  optionally uses OpenMP for faster execution on multiprocessor machines.
The documentation provided with the code includes sample plots.

\subsection{Main program}
\label{modulation}
The code package includes an example ($solar\_ic\_driver.cc$) containing the parameters editable by the users depending on their requirements. 
The parameters to be defined are: radius, temperature, distance of the Sun (star), energy range and grid factor for both CR electrons and positrons, gamma rays, grid spacing (linear or logarithmic) of the angle from the Sun, integration range and steps,  
all-electron spectra models including cases with free parameterizations.

\subsection{CR electrons and positrons}
\label{modulation}
In the public version many models for the electron and positron spectra are implemented to the $LeptonSpectrum$ class, including those based on existing publications \citep{Orlando2006, Orlando2008, Moskalenko, Abdo2011}. The model parameters are freely definable. 

Additional examples are reported in Section~\ref{sec3}.
Moreover, additional electron spectra and positron spectra can be defined by the user by adding named models to the same class. 
The CR modulation due to the solar wind is described with the force-field approximation, but this can be freely extended to more realistic models that we plan to include in the future.

\subsubsection{Solar (stellar) modulation treatment}
In StellarICS the solar modulation is treated depending on the model for CR electrons and positrons used. Models as the ones published in \citep{Orlando2008, Moskalenko, Abdo2011} implement the modulation formalism described in this section. The same or similar models can be applied to the stars.
The CR electron plus positron spectrum is given by the well known force-field approximation  used to obtain the modulated differential intensity $J(r,E_{e})$ at the kinetic energy of the particle $E_{e}$ and distance $r$ from the Sun (star) \citep{gleeson}. As demonstrated by  \cite{caballero} it is a good approximation for CRs in the heliosphere. The differential intensity of the modulated spectrum is given by:
\begin{equation}
J(r, E_{e})=~J ( \infty, ~E_{e}+ ~\Phi(r) ) ~ \frac{E_{e}~(E_{e}+2E_{0})}{(E_{e}+~\Phi(r)+2E_{0})(E_{e}+\Phi(r))} 
\label{eq1}
\end{equation}
where $E_{e}$ is kinetic energy of the particle, $E_{0}$ is the electron rest mass, $J ( \infty, ~E_{e}+ ~\Phi(r))$ is the
local interstellar all-electron spectrum, and
$\Phi(r)=(Ze/A)\phi$, where $\phi$ is the modulation potential, $Z$ the charge number, and $A$ the mass number.

In order to compute the modulation potential as a function of the distance from the Sun (and star), StellarICS implements various parameterizations. Based on \cite{fujii} from 100~AU to 1~AU, neglecting the time dependence and normalizing the spectrum at 1~AU,
for Cycles 20/22 previous works \cite{Moskalenko, Orlando2008, Abdo2011} used:

\noindent
\begin{equation}
\Phi_1(r)= \frac{\Phi_0}{1.88}\left\{
\begin{array}{ll}
r^{-0.4} - r_b^{-0.4}, & r\ge r_0,\\
0.24 + 8 (r^{-0.1} - r_0^{-0.1}), & r<r_0, 
\end{array}
\right.
\label{eq2}
\end{equation}

\noindent
where $\Phi_0$ is the modulation potential at 1~AU, $r_0=10$~AU, and $r_b=100$~AU is the heliospheric boundary. For Cycle 21 the same authors used:

\noindent 
\begin{equation}
\Phi_2(r)= \Phi_0 (r^{-0.1} - r_b^{-0.1}) / (1 - r_b^{-0.1}).
\label{eq3}
\end{equation}

\noindent
For $r \geq r_b$, we have that $\Phi_1(r)$ = $\Phi_2(r)$ = 0. \\
This treatment of the solar modulation is found to work in the outer solar system. 
Indeed, these formulae were derived for $r>  1$~AU. Closer to the Sun, at $r<1$~AU, 
CR transport is very uncertain. 
Because the application of eq.~\ref{eq2} and eq.~\ref{eq3}  in the inner solar system (i.e. $r<  1$~AU) produces a substantial suppression of CRs, further alternative models are implemented, which do not have additional solar modulation between Earth and Sun. These models provide an upper limit to the expected IC emission.
These implement a constant $\Phi(r)$ = $\Phi_0$ for the inner heliosphere, where the value is constrained by CR all-electron measurements. \\

\subsection{Solar and stellar radiation field}
The black-body formulation for the solar (and stellar) radiation field as a function of the distance from the Sun (and star) is calculated in the class called $StarPhotonField$. The formulation is characterized by the effective temperature of the photosphere following the Stephan-Boltzmann equation. Following our previous approach \cite{Orlando2008}, the distribution of photon density close to the Sun (star), as extended source where the simple inverse-square law is inappropriate, is given by integrating over the solid angle of the solar (stellar) surface (see Fig.~2 of \cite{Orlando2008} for details), resulting in: 
\begin{eqnarray}
n_{ph}(E_{ph},r)=0.5 ~n_{BB}(E_{ph}) \left[ 1-\sqrt{1-(R_\odot /r)^{2}}\right]
\label{eq4}
\end{eqnarray}
with $R_\odot$ radius of the Sun (star), $E_{ph}$ solar (stellar) photon energy, and $n_{BB}$ black-body photon density.
For large distances from the Sun, eq.~\ref{eq4} reduces to the usual inverse-square law.

\subsection{IC emissivity and differential cross-sections}
\label{modulation}
Cross-sections and IC emissivity are calculated in class $InverseCompton$. \\
The IC emissivity for a given all-electron spectrum and radiation field is computed following the formulation in \cite{Orlando2006}:

\begin{equation}
%\label{eq1}
\epsilon(E_{\gamma})=\int dE_{e} %\times
%\nonumber\\
%\times 
\int \sigma_{KN}(\gamma,E_{ph},E_{\gamma})~ n_{ph}(E_{ph},r)~c~N(E_{e},r)~dE_{ph}
\label{eq5}
\end{equation}
where $N(E_{e},r)$ is the electron density, with $E_{e}$ kinetic energy of the particle as defined above, $n_{ph}(E_{ph},r)$ is the solar (or stellar) photon density as function of the distance from the Sun (or star) and of the solar (stellar) photon field $E_{ph}$ as in eq.~\ref{eq4}, $\sigma_{KN}$ is the Klein-Nishina cross-section, $\gamma=E_{e}/m_{e}$, and $E_{\gamma}$ is the energy of the gamma-ray IC photons.
The CR all-electrons are assumed to have an isotropic angular distribution everywhere in the heliosphere, as is the case for CRs in the Galaxy.

Both isotropic and anisotropic Klein-Nishina cross-sections are included, as described in \cite{StrongThesis} and in \cite{Orlando2008}. 
In particular for the anisotropic cross section we have:

\def\epsone{E_{ph}}
\def\epstwo{E_{\gamma}}
\def\epsonep{E_{ph}^\prime}
\def\epstwop{E_{\gamma}^\prime}
\def\Ee{E_e}
\def\me{m_e}
\def\etap{\eta^\prime}

\begin{equation}
{d\sigma_{KN}\over d\epstwo}=\pi{r_e}^2
 {\me\over\epsonep\Ee}
[
({\me\over\epsonep})^2({v\over 1-v})^2 - 2{\me\over\epsonep}{v\over 1-v}
+ (1 - v) + {1\over 1 - v}
 ]
\label{eq5b}
\end{equation}

with $E_{ph}$ target photon energy in lab; $E_{\gamma}$ gamma-ray energy in lab, $E_{ph}^\prime$ target photon energy in electron system with $\epsonep=\gamma\epsone(1 - \cos\eta)$, $E_{\gamma}^\prime$ gamma-ray energy in electron system with $\epstwo=\gamma\epstwop(1 - \cos\etap)$, $E_e$ electron energy, $\eta$ scatter angle of photon in lab, and $\etap$ scatter angle of photon in electron system.

The Thompson cross-section formulation is also included for a cross-check. \\
For comparison with other formulations of the IC, the reader is referred to the
documentation in the StellarICS package (IC$\_$documentation.pdf).
In particular, in that documentation, the form in terms of the energy transfer $v=E_\gamma/E_e$,
as used in the code, is  
compared to other forms (e.g. as used in GALPROP \cite{Moskalenko2000, Strong2007, Orlando2013, Gulli}).

\subsection{Calculation of the emission: gamma-ray flux and intensity}
\label{calculation}
The calculation of the IC emission integrated along the line-of-sight is performed in the class $SolarIC$. 
This class uses the input parameters from $solar\_ic\_driver$, the electron and positron spectrum parametrization from $LeptonSpectrum$, the photon field of the Sun (or star) from $StarPhotonField$ as a black-body source, and the IC cross-sections based on the Klein-Nishina formula from $InverseCompton$ which can be isotropic or anisotropic. 

The IC intensity spectrum along the line-of-sight $s$ for a given all-electron spectrum and radiation field follows the formulation in \cite{Orlando2006, Orlando2008}: \\
\begin{equation}
\label{eq6}
I(E_{\gamma})=\frac{1}{4\pi}\int\epsilon(E_{\gamma}, s, \theta) ds
\end{equation}
where $ds$ is the infinitesimal increment along the line-of-sight and $\theta$ is the angular distance from the Sun (star); $s$ goes from 0 at the Earth, to the user's defined maximum distance along the line-of-sight. In the direction to the Sun (star) surface $s_{max} = d - R$, with $d$ distance to the Sun (star) and $R$ solar (stellar) radius.
The available option to use a logarithmic angular grid, instead of the default linear angular grid, is useful to resolve the rapid variation near the solar disk while still covering the full angular range required far from the Sun.

StellarICS also accounts for the following effect.
The solar (or stellar) radiation field is anisotropic since the photons
are traveling away from the surface,
and the radiation is radial for distances $r~\gg~R$. The IC reaction rate
depends on the relative velocity of electron and photon, $2c~(1-\cos \eta)$
where $\eta$ is the angle between electron and photon momenta.
This causes larger variation with
angle from the Sun (star) than would be the case for isotropic radiation. See
also the discussion of the dip in the direction of the disk in Section~\ref{UpdatedModels}.

\begin{table*}
\begin{center}
\caption{Extensions names and contents of the output FITS file and corresponding  units. 
%\hline
$I(E,\theta)$ is the differential intensity for gamma-ray energy $E$ at angle $\theta$ from the solar centre.}
%\hline
\begin{tabular}{llccl}
\hline
\hline
Extension's     &        Contents&        Row &    Column  & Units\\
Number and Name                  &                &  Variable  &   Variable &      \\
                   \hline
1.     Differential intensity\\
~~~~profile          &         $I( E_{\gamma}, \theta)$      &   $E_{\gamma}$     &       $\theta$   &       cm$^{-2}$ sr$^{-1}$ s$^{-1}$ MeV$^{-1}$\\
2. Energy integrated\\ 
~~~~profile                    &        $I(>E_{\gamma}, \theta)$     &      $E_{\gamma}$       &     $\theta$    &    cm$^{-2}$ sr$^{-1}$ s$^{-1}$ \\
3.  Angle integrated profile                     &       $I( E_{\gamma},<\theta)$     &     $E_{\gamma}$       &     $\theta$    &     cm$^{-2}$ s$^{-1}$ MeV$^{-1}$\\
4.  Energy and angle\\
~~~~integrated profile   &             $I(>E_{\gamma},<\theta)$     &  $E_{\gamma}$    &       $\theta$      &      cm$^{-2}$ s$^{-1}$\\
5.    Spectrum for angles                            &  $I( E_{\gamma},\theta)$     &      $\theta$   &     $E_{\gamma}$        &     cm$^{-2}$ sr$^{-1}$ s$^{-1}$ MeV$^{-1}$\\
6.    Spectrum times\\
~~~~Esquared for angles   &            $I( E_{\gamma},\theta) \times E_{\gamma}^{2}$  &     $\theta$ & $E_{\gamma}$ &     cm$^{-2}$ sr$^{-1}$ s$^{-1}$ MeV\\
7.   Spectrum for\\
~~~~integrated angles                 & $I( E_{\gamma},<\theta)$      &     $\theta$  &      $E_{\gamma}$       &     cm$^{-2}$ s$^{-1}$ MeV$^{-1}$\\
8.  Spectrum times\\
~~~~Esquared for integrated\\
~~~~angles & $I( E_{\gamma},<\theta) \times E_{\gamma}^{2}$   &    $\theta$ & $E_{\gamma}$ &    cm$^{-2}$ s$^{-1}$ MeV\\
9.     Energies                               &         $E_{\gamma}$&    $E_{\gamma}$              &                   &            MeV\\
10.     Angles                                        &       $\theta$  &  $\theta$       &                 &            degrees\\
\label{Table1}
%\hline
\end{tabular}
%\hline    %  causes tex error
%\hline
\end{center}
\end{table*}

\subsection{Output}
\label{sec26}
\label{output}
The output is a FITS file with multiple extensions for a convenient use. 
Extensions names and contents of the output FITS file and respectively units are listed in Table~\ref{Table1}.

In addition, the code outputs an $idl$ program which generates  spectra and profiles directly.

\section{Calculations for the quiet Sun from keV to TeV}
\label{sec3}
This section presents examples and calculations to the solar IC emission from keV to TeV energies
according to the available CR all-electron measurements under various solar conditions.

\subsection{Updated models for studies with Fermi-LAT}
\label{UpdatedModels}
Pamela \cite{Pamela} and AMS-02 \cite{AMS_ele} 
all-electron measurements show noticeably differences with respect to the measurements used in our previous works \cite{Orlando2006, Orlando2008, Abdo2011}. 

\begin{figure}.
\center
\includegraphics[width=0.7\textwidth]{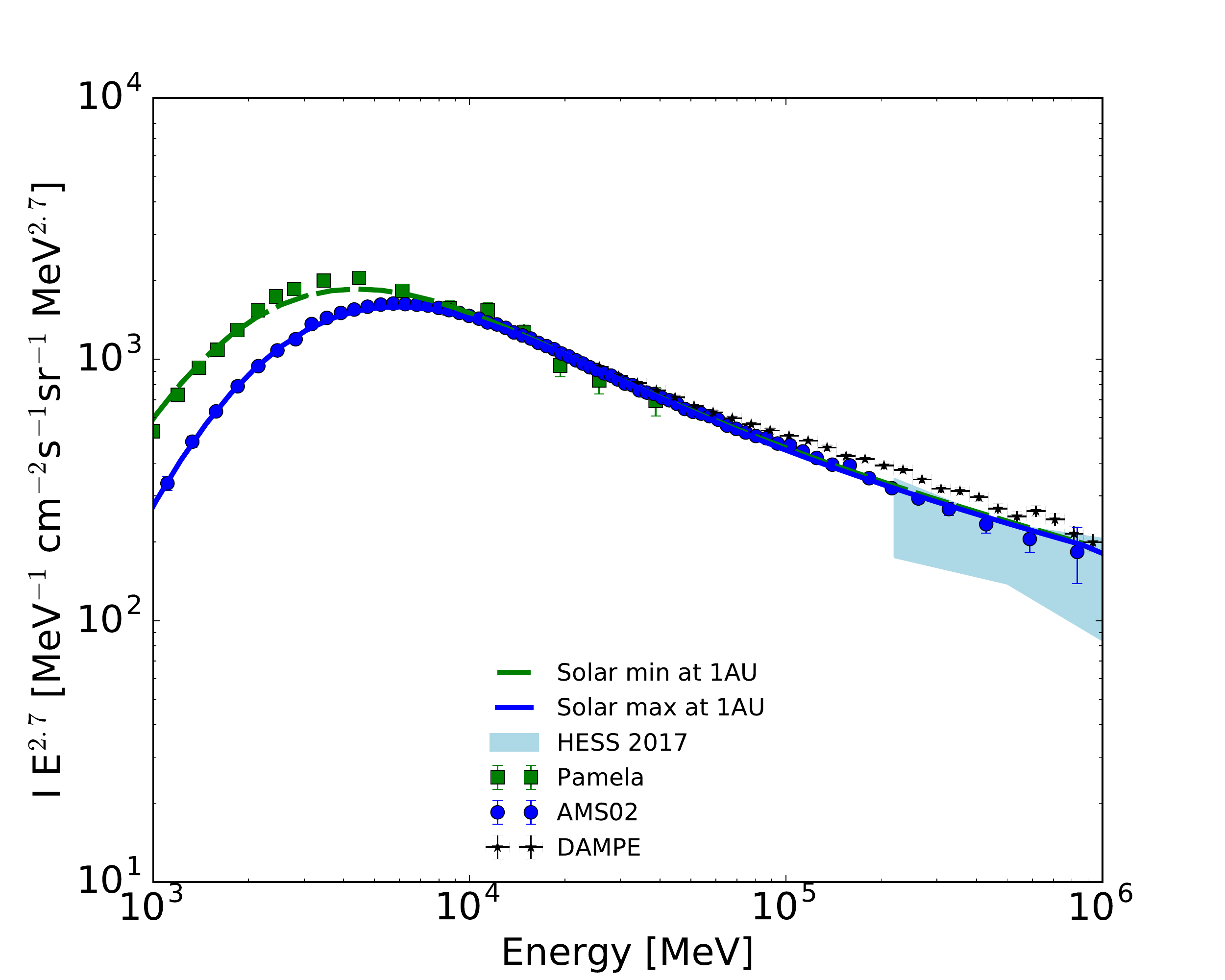}
\caption{\label{fig1} All-electron spectra for two sample models compared with data. The green dashed line shows the model mainly based on the Pamela \cite{Pamela} all-electron spectrum for the period of 2008 (green squares), representative of the solar minimum; the blue solid line shows the model mainly based on the AMS-02 \cite{AMS_ele} all-electron spectrum for the period of 2013 (blue points), representative of the solar maximum. Both modeled spectra are fitted to AMS-02 data above 40~GeV. DAMPE \cite{DAMPE} data (black crosses) and HESS \cite{HESS} preliminary data (cyan region) are also shown. 
}
\end{figure}

\begin{figure}[h]
\center
\includegraphics[width=0.49\textwidth]{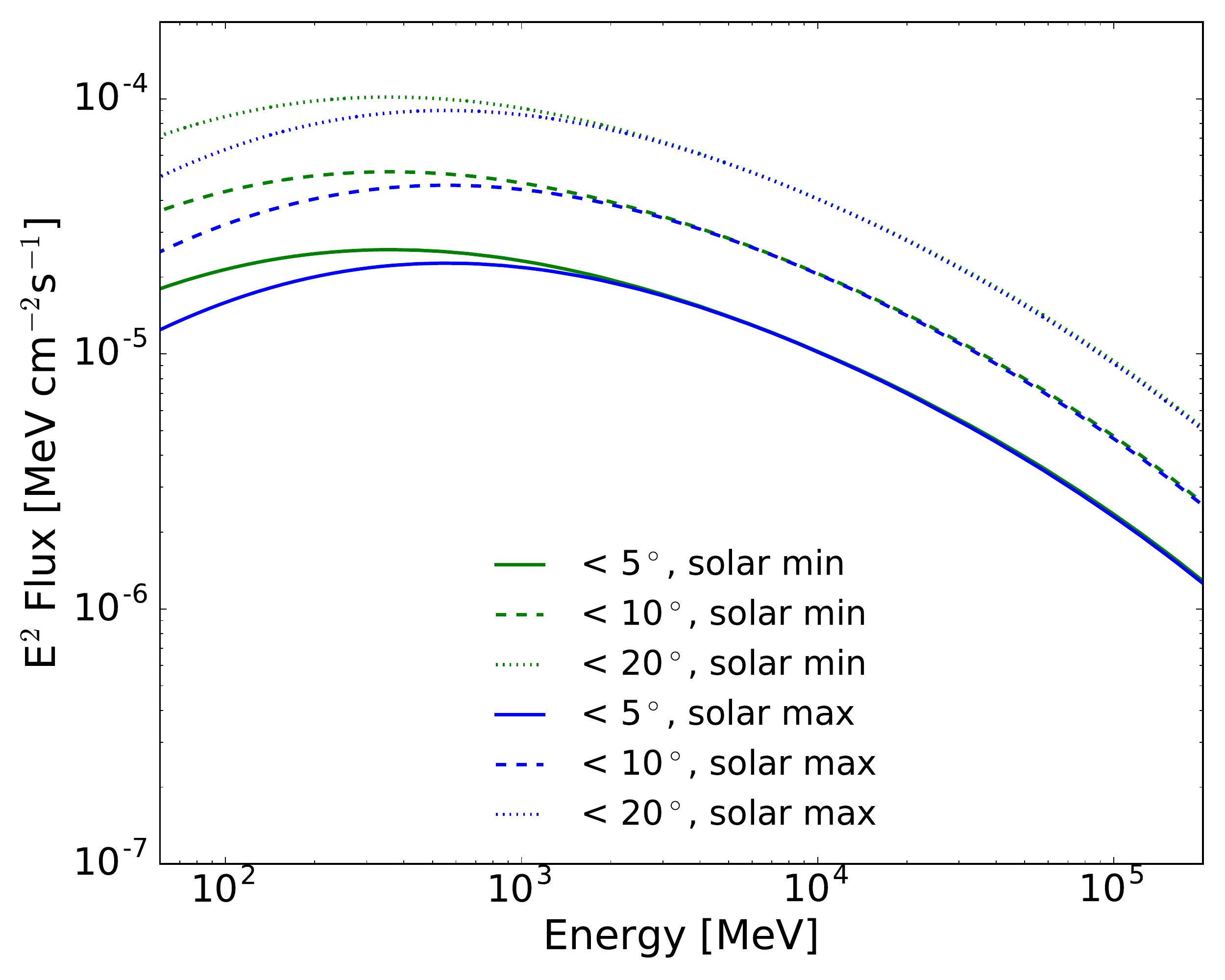}
\includegraphics[width=0.49\textwidth]{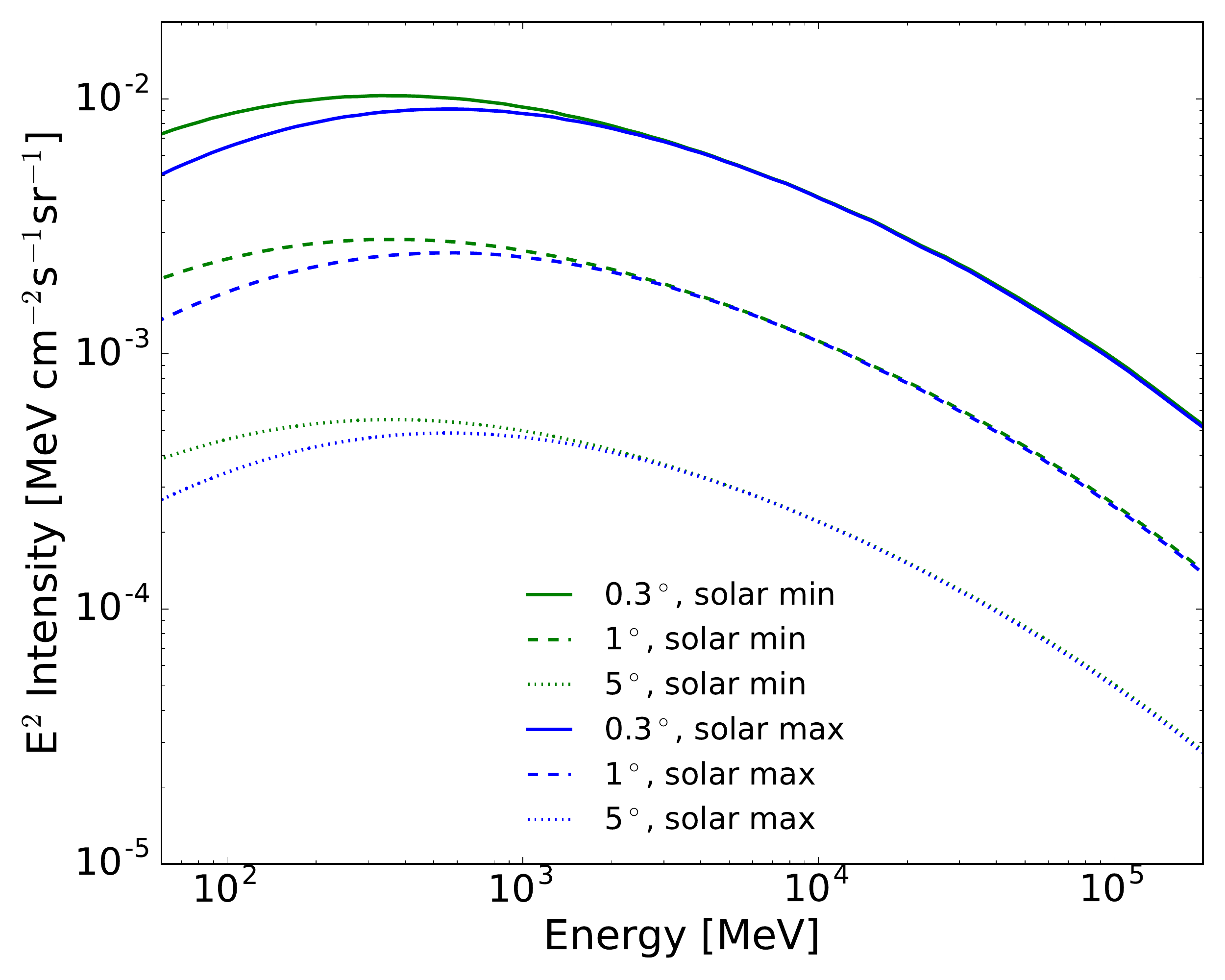}
\caption{\label{fig2}  {\it Left:} Calculated IC spectral flux in the energy range of Fermi-LAT integrated over areas with various angular amplitudes based on the green and blue all-electron spectra shown in Figure~\ref{fig1}. Solid lines are the spectral fluxes integrated within 5$^\circ$ of the Sun, dashed lines within 10$^\circ$ of the Sun, and dotted lines within  20$^\circ$ of the Sun. 
{\it Right:} Calculated IC spectral intensity in the energy range of Fermi-LAT for various angular distances from the Sun for the green and blue all-electron spectra shown in Figure~\ref{fig1}. Solid lines are the spectral intensity at 0.3$^\circ$ from the Sun, dashed lines at 1$^\circ$ from the Sun, and dotted lines at 5$^\circ$ from the Sun.
In both plots the blue lines represent the solar maximum condition (all-electron model that fits AMS-02 data), while the green lines represents the solar minimum condition (all-electron model which fits Pamela data).}
\end{figure}

\begin{figure}[h]
\center
\includegraphics[width=0.65\textwidth]{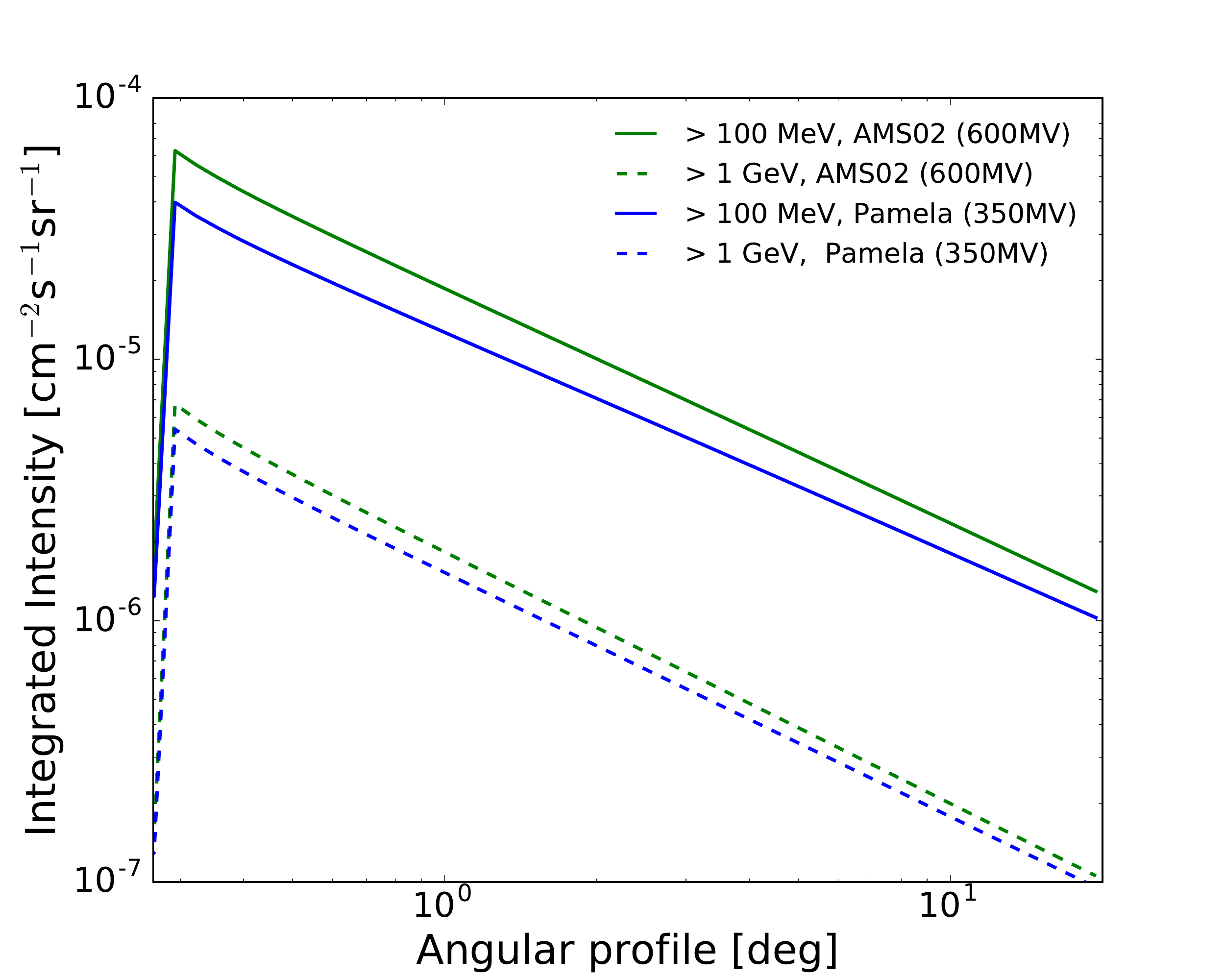}
\caption{\label{fig3} Calculated IC intensity profile integrated above 100 MeV (solid lines) and above 1 GeV (dashed lines) for the green and blue all-electron spectra shown in Figure~\ref{fig1}, with $\Phi(r)$ given by eq.\ref{eq3} for $r~\leq r_b$, i.e. also for $r~\leq$~1~AU. The blue lines represent the solar maximum condition, while the green lines represents the solar minimum condition.}
\end{figure}

Here we present some examples of IC models that incorporate the local interstellar all-electron spectra that fit Pamela and AMS-02 measurements at 1~AU.
In more detail, we define two baseline all-electron models: the first one mainly based on the Pamela all-electron spectrum for the period of 2008 \cite{Pamela}, which we consider the representative model for periods of solar minimum; the second one mainly based on the AMS-02 all-electron spectrum for the period of 2013 \cite{AMS_ele}, which we consider the representative model for periods of solar maximum. Both modeled spectra are fitted to AMS-02 data above 40 GeV, where AMS-02 data are more precise than Pamela and the solar modulation is not important.  As shown in \cite{Orlando2008}, all-electrons from ~1~GeV to ~1~TeV are responsible for generating the solar IC emission in the $\sim$30~MeV - 200~GeV Fermi-LAT energy range.
Figure~\ref{fig1} shows the all-electron spectra used to generate the two baseline IC models that can be used for the Fermi-LAT studies.   
The blue solid line and the green dashed line represent the models that fit the all-electron spectra at 1~AU respectively for AMS-02 and Pamela data. At higher energies the models can reproduce also recent HESS\footnote{https://www.mpi-hd.mpg.de/hfm/HESS/pages/home/som/2017/09/} preliminary data \cite{HESS}. DAMPE data \cite{DAMPE} are also plotted for comparison. 

As an example,  the plot on the left of Figure~\ref{fig2} shows the IC spectral flux integrated in 5$^\circ$ (solid lines), 10$^\circ$ (dashed lines), and 20$^\circ$ (dotted lines) around the Sun based on the two all-electron spectra of different solar conditions (green and blue spectra in Figure~\ref{fig1} respectively for solar minimum and maximum conditions). Spectral fluxes correspond to extension 8 in the output FITS file (see Table~\ref{Table1}). Here and throughout the paper we assume the temperature of the Sun to be 5778 K, the solar radius to be 6.955~$\times$10$^{10}$~cm, and the Sun's distance to be 1 AU.
The right plot in the same figure shows the spectral intensity (per steradian) for the same two all-electron spectra in Figure~\ref{fig1} for various angular distances from the Sun: 0.3$^\circ$ (solid lines), 1$^\circ$ (dashed lines), and 5$^\circ$ (dotted lines). Intensities correspond to those in extension table 6 in the output FITS file (see Table~\ref{Table1}). 
For illustration these two models in Figure~\ref{fig2} assume a constant modulation potential in the inner heliosphere, with the parameterization reported in eq.~\ref{eq3} for the outer heliosphere ($r$ > 1~AU); thus they do not assume any additional modulation in the inner heliosphere, i.e. $\Phi(r) = \Phi_{0}$ for $r$~<~1 AU in eq.\ref{eq3}, with $\Phi_{0} = 600~MV$ for solar maximum and $\Phi_{0} = 350~MV$ for solar minimum. 
For comparison the isotropic diffuse gamma-ray background (IGRB) has spectral intensity of (9.5~$\pm$~0.8) $\times$10$^{-4}$ MeV~cm$^{-2}$s$^{-1}$sr$^{-1}$ \cite{IGRB} at 100 MeV.

Alternatively, we can  extend eq.\ref{eq3} also to the inner heliosphere, when $\Phi(r)$ is given by eq.\ref{eq3} for any $r~\leq r_b$ for the same all-electron AMS02 and Pamela spectra in Figure~\ref{fig1}. 
The angular profiles of the two IC models (extension 2 in the output FITS file as in Table~\ref{Table1}) for these two cases are reported in Figure~\ref{fig3}. These profiles are integrated in energies above 100~MeV and above 1~GeV for illustration. Note that the effect of solar modulation is smaller for $>$ 1~GeV than for $>$ 100~MeV, as expected. 
Emission from the disk extension between the Earth and the Sun, which is not occulted by the disk itself, is expected.
However, the emission within the disk extension is substantially reduced. This is due to the Klein-Nishina formulation that decreases in the disk, going to 0 at $\theta = 0$.

It is instructive to compare the IC emission with well-known reference levels:
for example the IGRB integrated above 100 MeV has a total intensity of (7.2~$\pm$~0.6) $\times$10$^{-6}$ cm$^{-2}$s$^{-1}$sr$^{-1}$ \cite{IGRB}, which can be below the solar IC  even up to 25$^\circ$~--~40$^\circ$ angular distance from the Sun, depending on the models. \\

\subsection{Calculations at keV and MeV energies}

\begin{figure}[h]
\center
\includegraphics[width=0.65\textwidth]{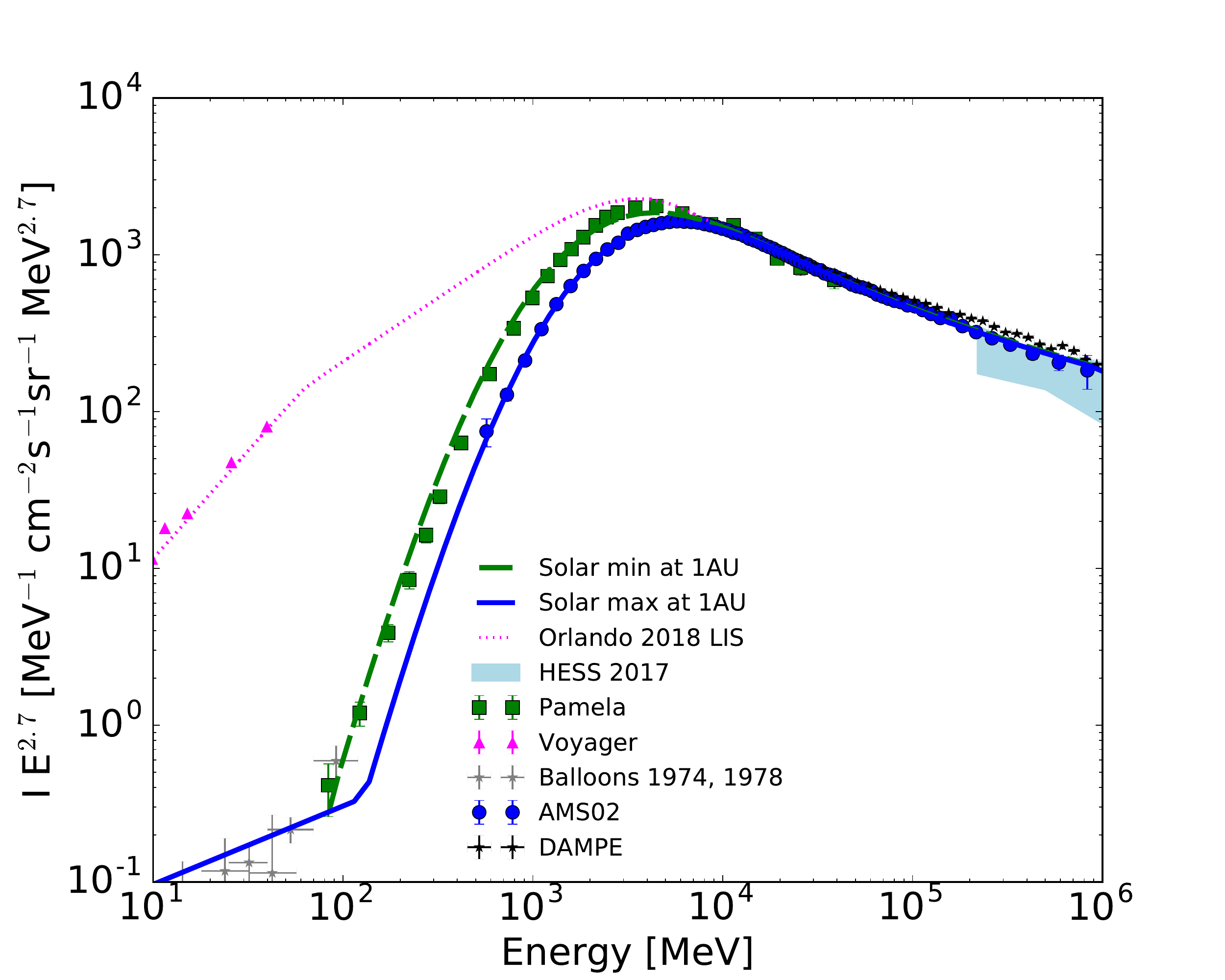}
\caption{\label{fig5} All-electron spectra for the two models compared with data. Green dashed line: model mainly based on the Pamela \cite{Pamela} all-electron spectrum for the period of 2008 (representative of the solar minimum); blue solid line: model mainly based on the AMS-02 \cite{AMS_ele} all-electron spectrum for the period of 2013 (representative of the solar maximum). Both are fitted to Balloons data from 1968 \cite{balloons1} and 1974 \cite{balloons2} of the interplanetary all-electrons. Both models are also fitted to AMS-02 data above 40 GeV. Also shown are DAMPE \cite{DAMPE} and HESS \cite{HESS} data and the interstellar Voyager \cite{Voyager} measurements. The magenta dotted line shows the local interstellar spectrum from \cite{Orlando2018} constrained by radio-microwave and gamma rays observations of the interstellar diffuse emission and direct CR measurements.}
\end{figure}

\begin{figure}[h]
\center
\includegraphics[width=0.49\textwidth]{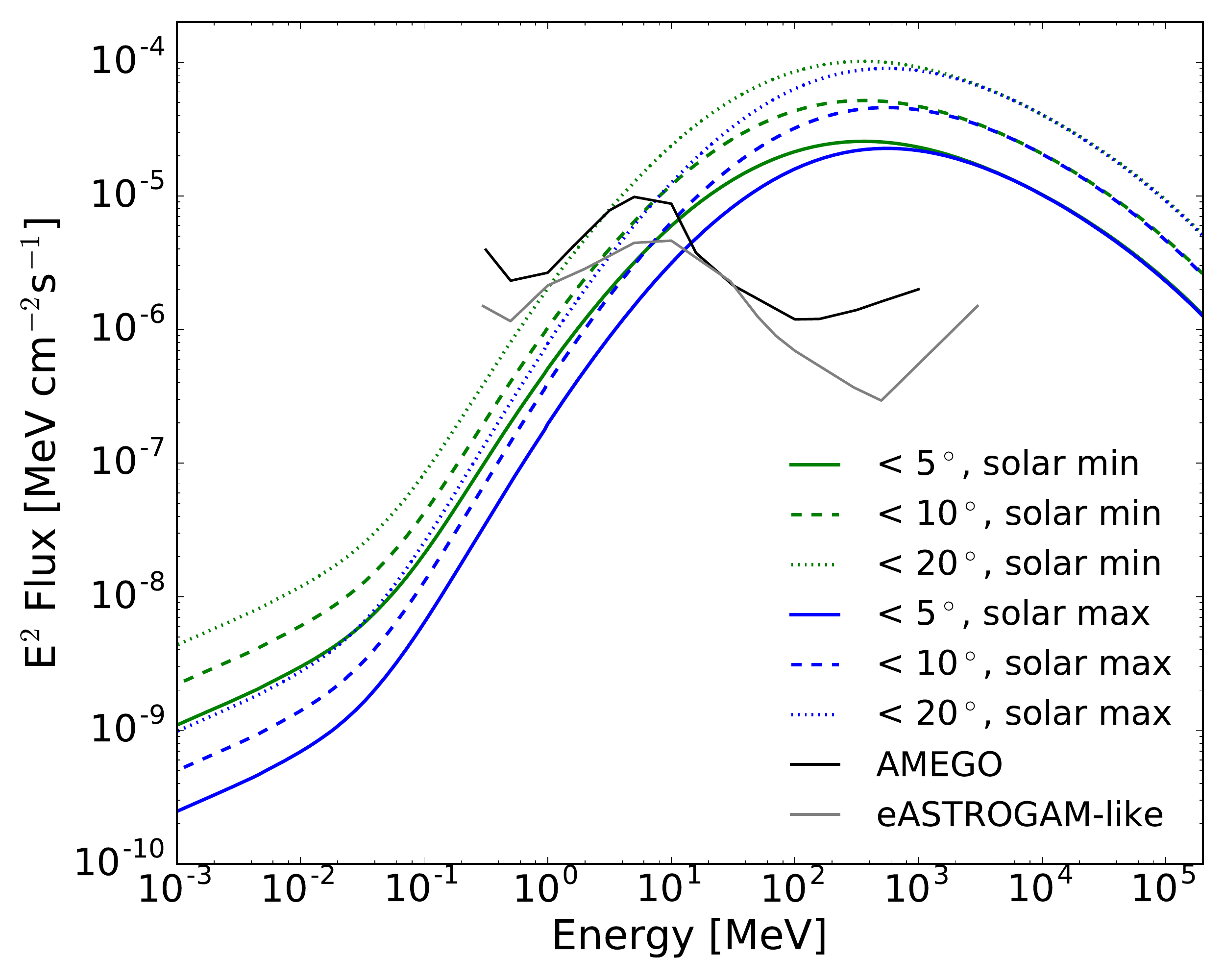}
\includegraphics[width=0.49\textwidth]{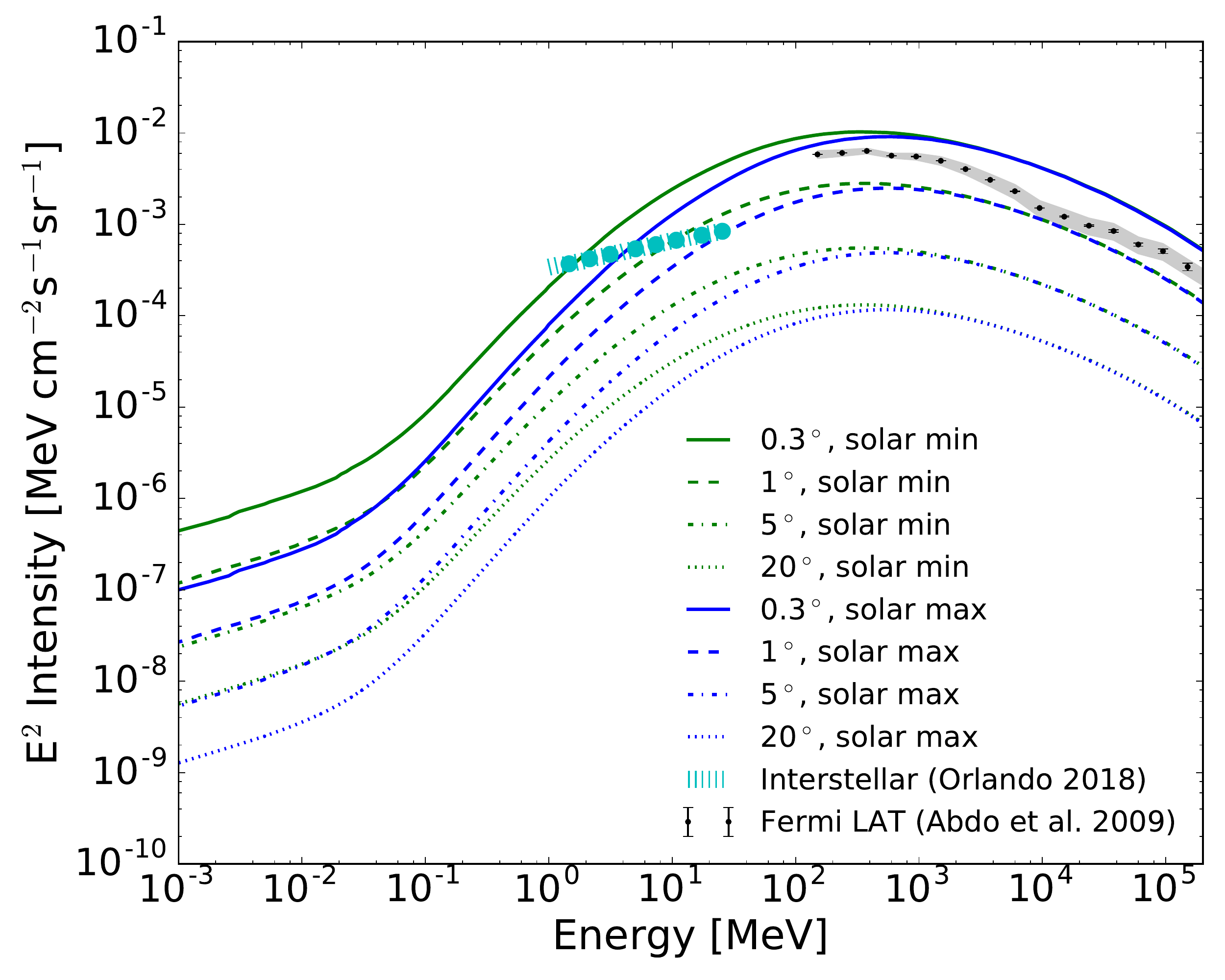}
\caption{\label{fig6} Extension of model calculations as in Figure~\ref{fig2} of the spectral flux (left) and spectral intensity (right) of the solar IC emission down to 1 keV. {\it Left:} Calculated IC spectral flux integrated over areas with various angular amplitudes based on the green and blue all-electron spectra shown in Figure~\ref{fig5}. Solid lines are the spectral fluxes integrated in 5$^\circ$ around the Sun,  while dashed lines in 10$^\circ$ around the Sun, and dotted lines in 20$^\circ$ around the Sun. AMEGO sensitivity (grey solid line, preliminary point source sensitivity for 5-year mission, private communication) and e-ASTROGAM-like instrument point source sensitivity \cite{eastrogam} (grey dashed line, for 1-year effective exposure) are also shown. 
{\it Right:} Calculated IC spectral intensity for various angular distances from the Sun based on the green and blue all-electron spectra shown in Figure~\ref{fig5}. Solid lines are the spectral intensity at 0.3$^\circ$ from the Sun,  while dashed lines at 1$^\circ$ from the Sun, dotted-dashed lines at 5$^\circ$ from the Sun, and dotted lines at 20$^\circ$ from the Sun.
In both plots the blue lines represent the solar maximum condition (all-electron model that fits AMS-02 data), while the green lines represents the solar minimum condition (all-electron model that fits Pamela data). Black points and grey region are data and systematic errors of the total emission at intermediate latitudes observed by Fermi-LAT \cite{diffuse1}, which includes the extended interstellar emission, the extragalactic component, and sources. The cyan region shows the interstellar emission at intermediate latitudes as predicted in \cite{Orlando2018}.}
\end{figure}  

We expect the Sun to emit IC gamma rays also at keV and MeV energies. 
KeV and MeV energies are of interest for various proposed missions, such as AMEGO, GECCO, and an e-ASTROGAM-like instrument.
AMEGO is designed to detect photons from 200 keV to 10 GeV, while GECCO from 50 KeV to 10 MeV, and e-ASTROGAM from 300 keV to 3 GeV. Moreover, this could be of interest of current and future X-ray telescopes looking at the Sun or even detecting the diffuse emission at larger solar distances (e.g. eRosita). 
Therefore, this section presents our IC calculations down to keV energies. 

Figure~\ref{fig5} shows the extended all-electron spectra below 1 GeV, which would be responsible for the IC emission at keV and MeV energies. Also in this case we show the two baseline IC models: the first one mainly based on the Pamela all-electron spectrum for the period of 2008, which we consider the representative model for periods of solar minimum (green line); the second one mainly based on the AMS-02 all-electron spectrum for the period of 2013, which we consider the representative model for periods of solar maximum (blue line). Both modeled spectra are fitted to available low-energy Balloon data from 1968 \cite{balloons1} and 1974 \cite{balloons2}. These are interplanetary all-electrons. The same figure also shows Voyager I all-electron measurements \cite{Voyager} in the interstellar space and a model from the local interstellar all-electron spectrum \citep{Orlando2018} that fits also Voyager I data.\\
The plot on the left of Figure~\ref{fig6} shows calculations for the IC spectral flux integrated in 5$^\circ$,  10$^\circ$, and 20$^\circ$ around the Sun based on the two all-electron spectra of different solar conditions: solar minimum (green lines) and solar maximum (blue lines) that correspond to the all-electron spectra shown in Figure~\ref{fig5} with the same color coding. Spectral fluxes correspond to extension~8 in the output FITS file (see Table~\ref{Table1}). AMEGO sensitivity (grey solid line, preliminary point source sensitivity for 5-year mission, private communication) and e-ASTROGAM-like instrument point source sensitivity \cite{eastrogam} (grey dashed line, for 1-year effective exposure) are also shown. Spectral sensitivities are for a point source, which would be different from the sensitivity for an extended source. 
If we want to directly compare models of extended emissions with the instrumental sensitivity, this should be given for extended emission where the angular resolution of the instrument would be taken into account and instrumental simulations would be performed, which is beyond the present effort and our knowledges of the specific instruments. For a more informative comparison, AMEGO spectral sensitivity is 4$\times$10$^{-6}$ MeV cm$^{-2}$s$^{-1}$ at 1~MeV, 4.8$\times$10$^{-6}$ MeV cm$^{-2}$s$^{-1}$ at 10~MeV, and 1$\times$10$^{-6}$ MeV cm$^{-2}$s$^{-1}$ at 100~MeV, with an angular resolution of 3$^\circ$ at 1~MeV and 10$^\circ$ at 10~MeV \footnote{https://asd.gsfc.nasa.gov/amego/technical.html}. Below 10 MeV it is expected that GECCO has even better sensitivity being a Compton telescope down to 100 keV, even though it may not be suitable to directly point to the Sun. 
As reported in \cite{eastrogam} an e-ASTROGAM-like instrument would have a sensitivity better than the following fluxes: 2$\times$10$^{-5}$ MeV cm$^{-2}$s$^{-1}$, 5$\times$10$^{-5}$ MeV cm$^{-2}$s$^{-1}$, and 3$\times$10$^{-6}$ MeV cm$^{-2}$s$^{-1}$, at 1~MeV, 10~MeV, and 500~MeV respectively for a 1-year effective exposure of a high Galactic latitude source.
%The same paper reports also the expected spectral extended-source sensitivity for 1 year of observations based on simulations for the inner Galaxy. This is of the order of a few 10$^{-5}$ MeV cm$^{-2}$s$^{-1}$sr$^{-1}$ below a few MeV, increasing to $\sim$10$^{-5}$ MeV cm$^{-2}$s$^{-1}$sr$^{-1}$ around 10 MeV, and decreasing again to a few 10$^{-5}$ MeV cm$^{-2}$s$^{-1}$sr$^{-1}$ above 30 MeV \cite{eastrogam}. 

The right plot in the same figure shows calculations for the spectral intensity (per steradians) based on the same two all-electron spectra in Figure~\ref{fig5} for various angular distances from the Sun: 0.3$^\circ$ (solid lines), 1$^\circ$ (dashed lines), 5$^\circ$ (dashed-dotted lines), and 20$^\circ$ (dotted lines). Intensities correspond to extension 6 in the output FITS file (see Table~\ref{Table1}). For comparison the expected interstellar emission at intermediate latitudes (i.e. 10$^\circ$$<|b|<$20$^\circ$) from \cite{Orlando2018} is also plotted, together with the data of the total emission detected by Fermi-LAT at intermediate latitudes by \cite{diffuse1}, which includes the interstellar emission as the main component, and the extragalactic background and sources as minor components.
Even though the expected solar IC emission above 100~MeV is comparable with the interstellar emission detected by Fermi LAT that acts as a confusing background, the solar IC component is definitively detected by Fermi LAT, and it has been currently studied in detail. As a consequence, because also below 100~MeV the expected solar IC emission is comparable with the expected interstellar emission, we expect the IC emission to be detectable by any instrument sensitive enough to diffuse emissions in this energy range. \\
This makes the solar IC emission very important for such proposed MeV missions as it is for Fermi-LAT.
On the contrary, the extragalactic X-ray background is $\sim$ 10$^{-2}$ MeV cm$^{-2}$s$^{-1}$sr$^{-1}$ at 1~keV and $\sim$ 3$\times$ 10$^{-2}$ MeV cm$^{-2}$s$^{-1}$sr$^{-1}$ at 10~keV making the solar large-scale emission hardly detectable below 100~keV (and hence for eRosita).\\
For illustration, the models in Figure~\ref{fig6} assume no additional modulation from Earth to the Sun, which gives an upper limit to the IC emission produced by Galactic and interplanetary CR all-electrons. However, various modulation models can be assumed to test gamma-ray data at MeV when they will be collected by AMEGO, an e-ASTROGAM-like instrument, or GECCO. Hence, such data will enable us to study the low-energy part of the all-electron spectrum in close proximity of the Sun.
As for the Fermi-LAT, given the broad distribution of the IC emission on the sky, the solar contribution to the diffuse MeV background cannot be neglected.

\subsection{Calculations at TeV energies}
Investigating possible TeV emission from the quiet Sun has recently sparked interest, especially because of the HAWC, ARGO, LHAASO experiments and the forthcoming next generation SWGO \cite{SWGO}. In fact, upper limits to the emission from the solar disk have been recently obtained with HAWC \cite{HAWC} and ARGO-YBJ \cite{Bartoli}. Hence, in this section we extend our calculations of the IC emission to the TeV energy band. 

\begin{figure}
\center
\includegraphics[width=0.65\textwidth]{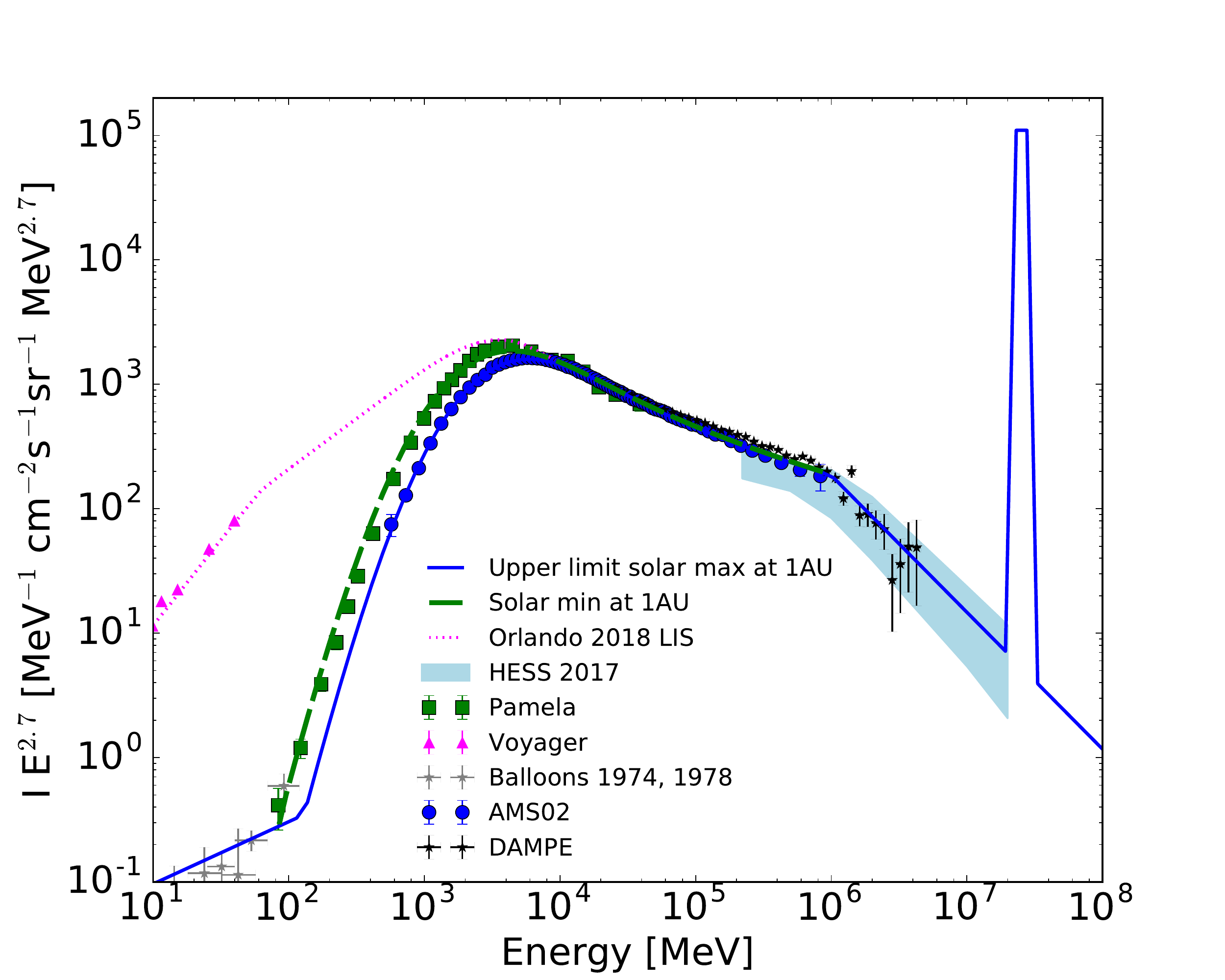}
\caption{\label{fig7} Modeled all-electron spectra compared with data. For energies below 1 TeV, details are provided in Figure~\ref{fig5}. Up to 20 TeV the modeled spectrum (blue line) reproduces the HESS data \cite{HESS} and other CR electron data, while at higher energies the upper limit to the spectrum (spike in the figure) is given by the CR proton spectrum and by ground-based observations of nearby sources. This spectrum represents an extreme case that provides the upper limit to the calculated IC emission.}
\end{figure}

\begin{figure}
\center
\includegraphics[width=0.49\textwidth]{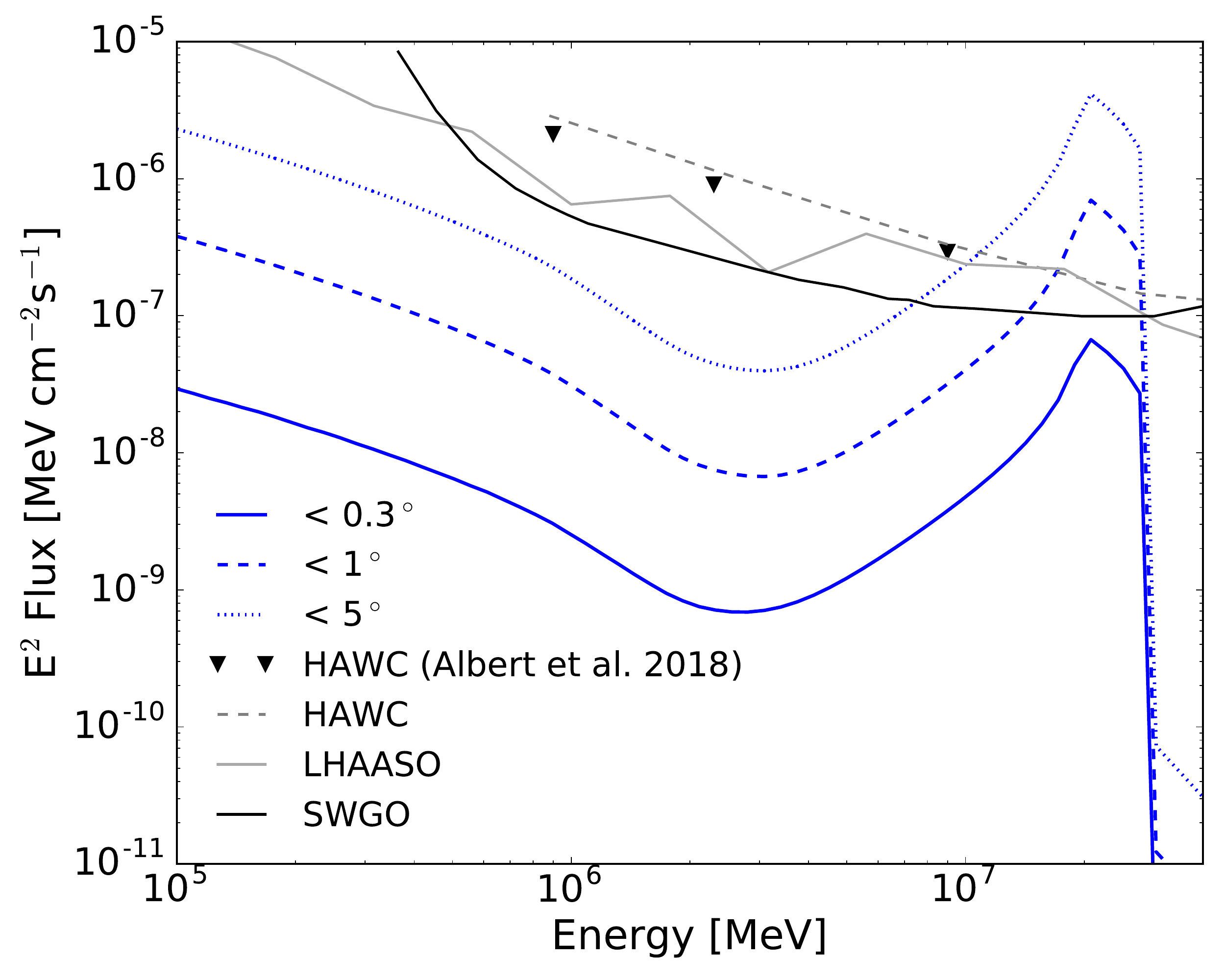}
\includegraphics[width=0.49\textwidth]{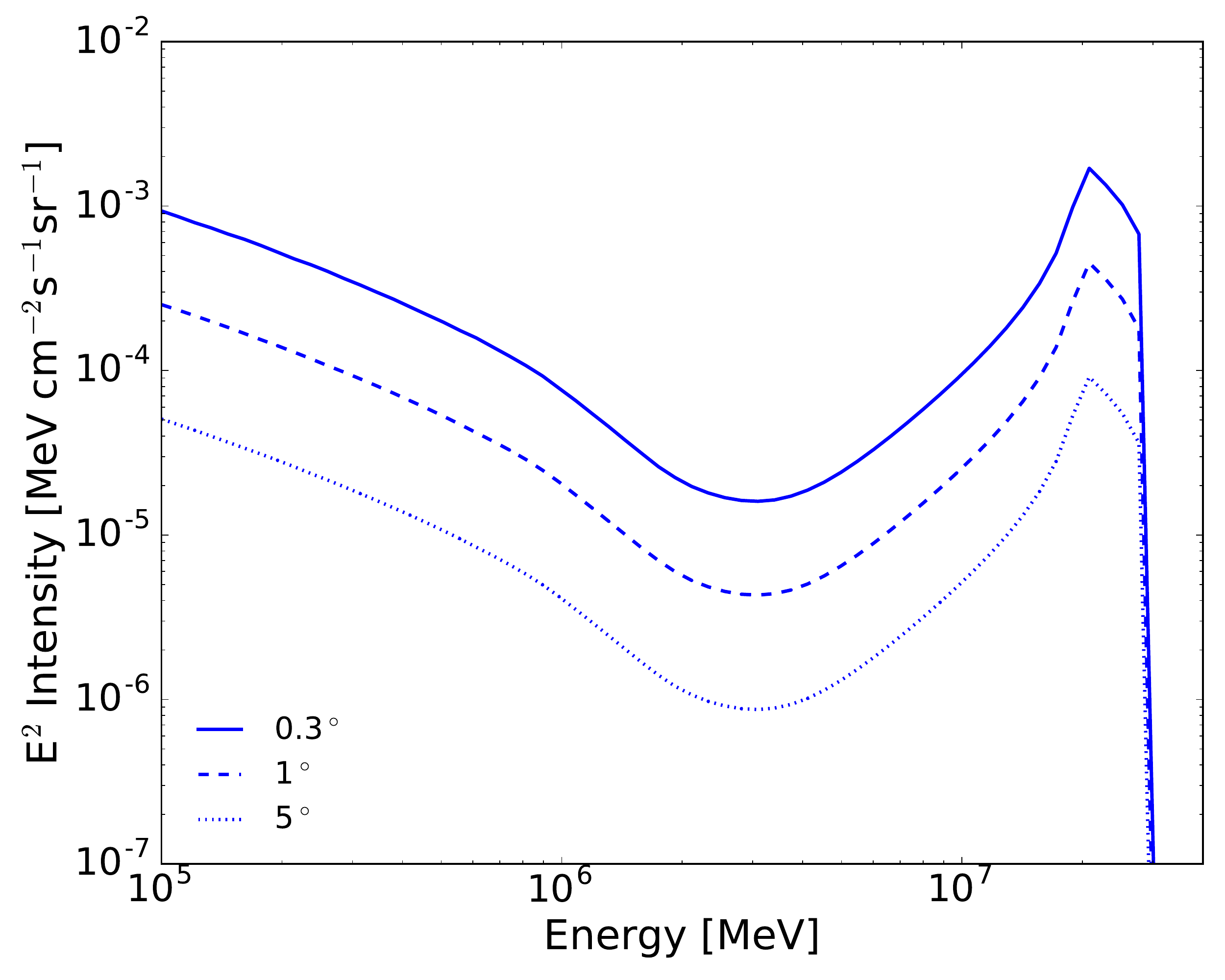}
\caption{\label{fig8} Extension of the solar IC model calculations at TeV energies based on the all-electron model as in Figure~\ref{fig7} (blue line). This gives the expected upper limit to the IC emission. {\it Left:} calculated spectral flux integrated over areas with various angular amplitudes: 0.3$^\circ$ (solid line), 1$^\circ$ (dashed line) and 5$^\circ$ (dotted line) around the Sun. Black triangles show HAWC upper limits of the disk emission \cite{HAWC}. HAWC 3-year sensitivity \cite{HAWC}(dashed grey line), LHAASO sensitivity \cite{LHAASO} (solid grey line, 1-year sensitivity private communications), and SWGO sensitivity \cite{SWGO} (solid black line, the SWGO {\it straw man} model) for point sources are also shown. {\it Right:} calculated spectral intensity for various angular amplitudes: 0.3$^\circ$ (solid line), 1$^\circ$ (dashed line) and 5$^\circ$ (dotted line) from the Sun. }
\end{figure}  

At TeV energy ranges the Klien-Nishina formulation would suppress the IC emission making the IC emission hardly observable by present and future telescopes. Hence in this example we calculate an upper limit to the expected IC emission for the case of the most extreme all-electron spectrum expected. In more details, Figure~\ref{fig7} shows an extreme all-electron spectrum that extends to 100~TeV (blue line). For energies below 1~TeV, the all-electron spectrum is constrained by all-electron CR measurements, whose details are provided in Figure~\ref{fig5}. Recent HESS data \cite{HESS} provide additional constraints to all-electrons up to 20~TeV. Because above 20~TeV there are no CR all-electron measurements, but there are instead CR proton measurements, we account for an extreme all-electron flux following the work in \cite{HAWC, Zhou}. In this extreme all-electron spectrum the upper limit to the flux is given by the CR proton flux and by gamma-ray observations of nearby sources. The consequence is that at $\sim$20~TeV this extreme model provides a spike, whose upper limit is given by the CR proton flux by assuming that CR all-electrons flux cannot be larger than the CR proton flux (usual all-electron to proton ratio is 1\%). At even higher energies the upper limit to the CR all-electron flux is instead given by ground-based arrays observations of nearby sources (for example as reported in \cite{Kistler, Kashiyama}). 
As in \cite{HAWC, Zhou} this represents an extreme case and it would give an upper limit to the all-electron spectrum, and hence to the calculated IC emission at TeV energies. 
 We are not investigating here the possible origin of such an extreme all-electron flux at these energies, if given by sources or dark matter (e.g. \cite{Profumo2005, Profumo2012, Huang, Coogan}). Above 100 TeV, we assume a sharp cut-off with no all-electrons.
%The assumed all-electron spectrum is in agreement with limits by ground-based arrays observations of nearby sources (for example as reported in \cite{Kistler, Kashiyama}). 

The resulting calculations  of the IC emission at TeV energies for this all-electron spectrum are presented in Figure~\ref{fig8}. 
The plot on the left of the figure shows calculations for the IC spectral flux integrated in 0.3$^\circ$ (solid line), 1$^\circ$ (dashed line) and 5$^\circ$ (dotted line) around the Sun. Spectral fluxes correspond to extension 8 in the output FITS file (see Table~\ref{Table1}). The same figure reports also recent upper limits from observations of the Sun with HAWC \cite{HAWC}. HAWC 3-year sensitivity (dashed grey line, \cite{HAWC}), LHAASO sensitivity (solid grey line, \cite{LHAASO}), and SWGO sensitivity (solid black line, \cite{SWGO}) for point sources are also shown. Also in this case, there may not be direct comparison of point-source sensitivities with calculations of the extended emission, because spectral sensitivities are for a point source, which may differ from the sensitivity for an extended source. For a more informative comparison by accounting for the angular resolution of the instruments HAWC spectral sensitivity for a point source for 5-year observations is about 4~$\times$~10$^{-6}$ MeV cm$^{-2}$s$^{-1}$ at 1~TeV, about 4~$\times$~10$^{-7}$ MeV cm$^{-2}$s$^{-1}$ at 10~TeV, and about 2~$\times$~10$^{-7}$ MeV cm$^{-2}$s$^{-1}$ at 30~TeV, with an angular resolution of 0.5$^\circ$ at 1~TeV improving at higher energies\footnote{https://www.hawc-observatory.org/details/xtreme.php}. LHAASO sensitivity for one year is 9~$\times$~10$^{-7}$ MeV cm$^{-2}$s$^{-1}$ at 1~TeV, about 3~$\times$~10$^{-7}$ MeV cm$^{-2}$s$^{-1}$ at 10~TeV, and about 10$^{-7}$ MeV cm$^{-2}$s$^{-1}$ at 30~TeV \cite{LHAASO}. The 5-year spectral sensitivity of SWGO is 4~$\times$~10$^{-7}$ MeV cm$^{-2}$s$^{-1}$ at 1~TeV, 5~$\times$~10$^{-8}$ MeV cm$^{-2}$s$^{-1}$ at 10~TeV, and about 3~$\times$~10$^{-8}$ MeV cm$^{-2}$s$^{-1}$ at 30~TeV with an angular resolution of 0.25$^\circ$ at 1~TeV strongly improving up to 0.1$^\circ$ at higher energies\footnote{https://www.swgo.org}. \\
The consequence is that the solar IC emission will be barely observable even by most sensitive planned telescopes at TeV and for an extreme all-electron spectrum. \\
The right plot in the same figure shows calculations for the spectral intensity (per steradian) for 0.3$^\circ$ (solid line), 1$^\circ$ (dashed line) and 5$^\circ$ (dotted line) angular distances from the Sun. This corresponds to extension table 6 in the output FITS file (see Table~\ref{Table1}).\\

\section{Examples of calculations for single stars}
Similarly to the Sun, other stars are expected to produce IC emission. This IC emission has been estimated to be detectable by Fermi-LAT from the closest and most luminous stars \cite{Orlando2006, OrlandoThesis}. Indeed, we expect the smooth large-scale interstellar IC emission to be clumpy towards these sources. Our ongoing effort \cite{DeMenezes} is dedicated to analyze single stars with Fermi-LAT data. \\ 
The IC emission was also accounted for the interpretation of freshly accelerated CRs in the Cygnus OB2 region \cite{cygnus}, where a model of IC emission from 1700 O and B stars based on \cite{OrlandoProc} was included in the analysis.

\begin{figure}[h!]
\center
\includegraphics[width=0.49\textwidth]{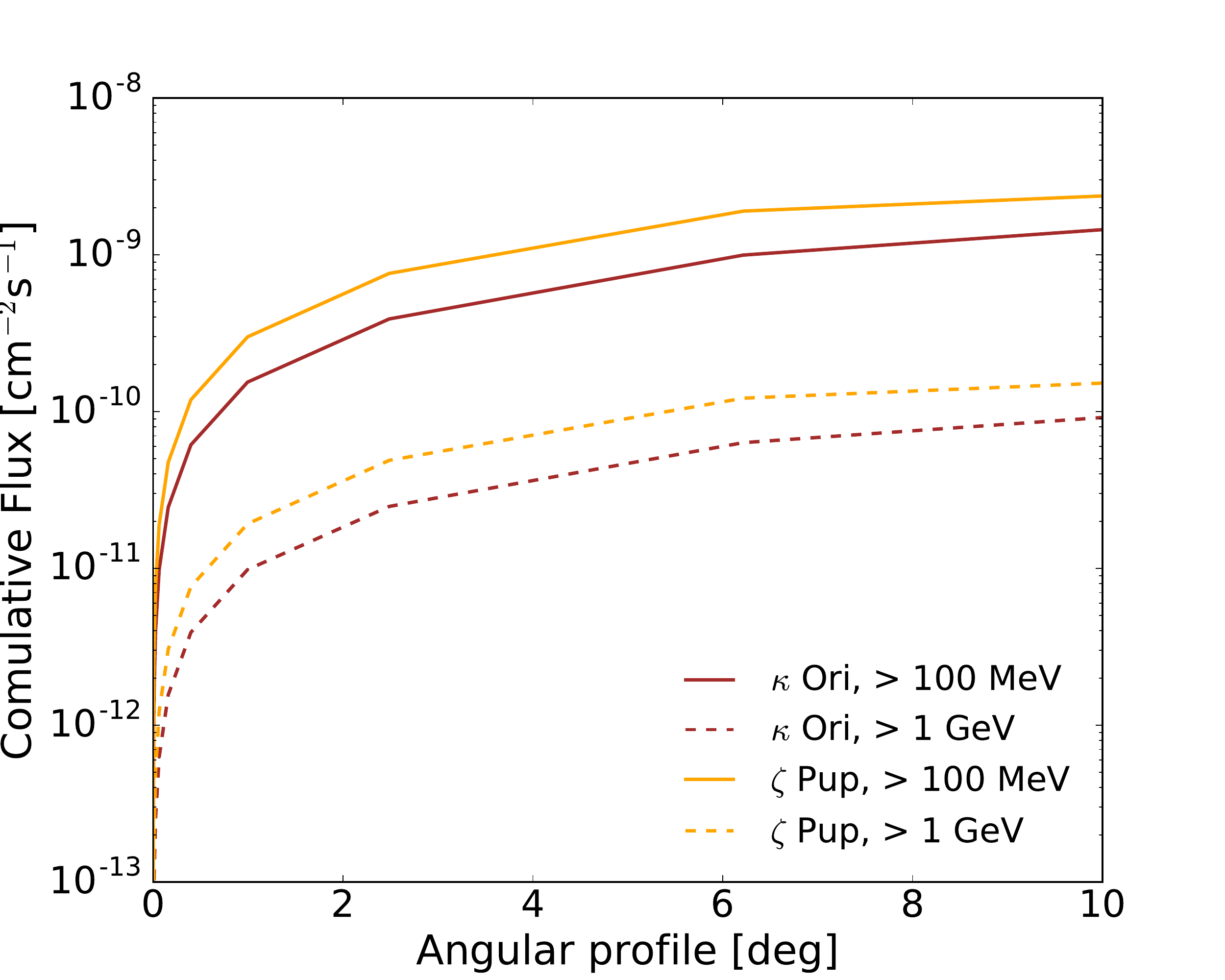}
\includegraphics[width=0.49\textwidth]{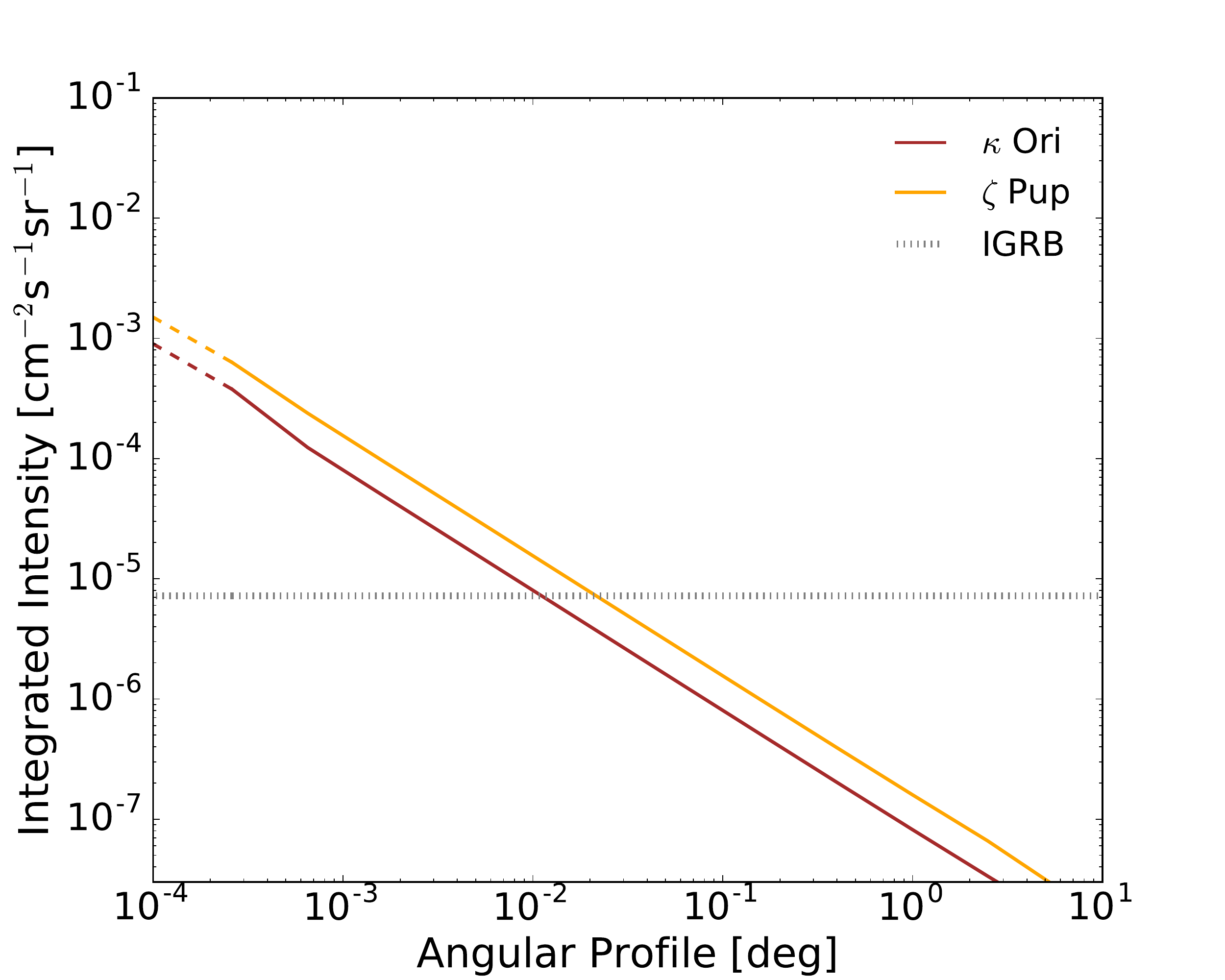}
\caption{\label{fig9} {\it Left:} Cumulative flux from $\kappa$~Ori (brown lines) and $\zeta$~Pup (orange lines) as a function of integration angle for integrated energies above~100 MeV (solid lines) and 1~GeV (dashed lines). {\it Right:} Intensity (per steradian) integrated above 100~MeV from $\kappa$~Ori (brown line) and $\zeta$~Pup (orange line) as a function of angular distance from the stars. The IGRB \cite{IGRB}, with a total intensity of (7.2$\pm$0.6) $\times$10$^{-6}$ cm$^{-2}$s$^{-1}$sr$^{-1}$, is shown for comparison (grey region).}
\end{figure}  

\begin{figure}
\center
\includegraphics[width=0.49\textwidth]{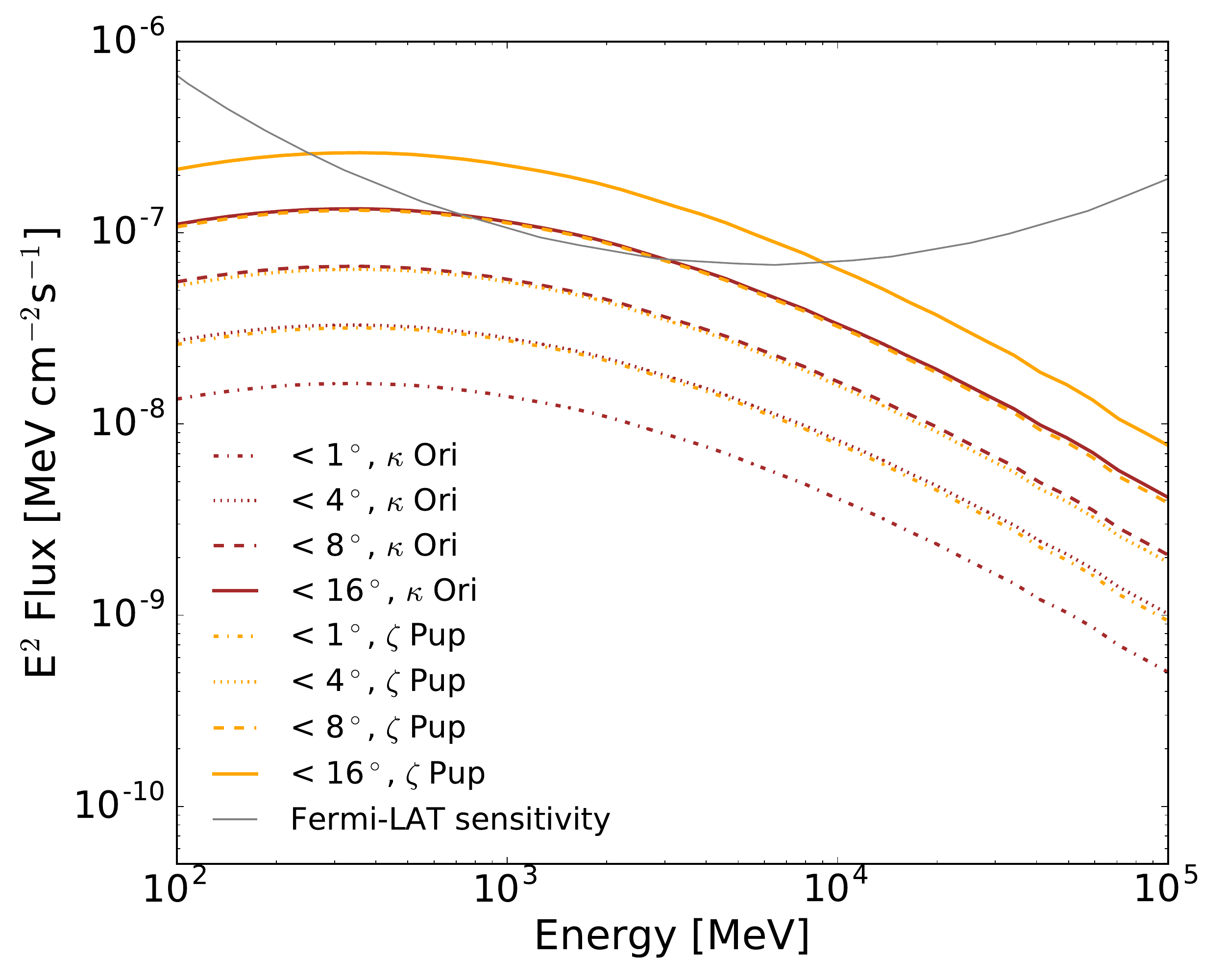}
\caption{\label{fig10} Calculated IC spectral flux of $\kappa$~Ori (brown lines) and $\zeta$~Pup (orange lines) in the energy range of Fermi-LAT integrated over areas with various angular amplitudes for the AMS02 spectrum with $\Phi_{0}$ = 600 MV in eq.~\ref{eq3}. Solid lines are the spectral fluxes integrated in 8$^\circ$ around the Sun, dashed lines in 4$^\circ$ around the Sun, and dotted lines in 1$^\circ$ around the Sun. The grey line is the Fermi-LAT point source sensitivity for 10 years}.
\end{figure}  

As discussed in previous sections, the formalism for the calculation of the IC emission from stars and Sun is similar. This section reports examples of IC emission from $\kappa$~Ori and $\zeta$~Pup. Further candidates are listed in \cite{Orlando2006}, \cite{OrlandoThesis}, and \cite{DeMenezes}. For $\kappa$~Ori, B0.5 Ia star, we use the following parameters: distance = 198~$pc$, temperature = 26500 $K$, and radius = 22.2 $R_\odot$. For $\zeta$~Pup, O4If(n)p star, we use the following parameters: distance = 330~$pc$, temperature = 40000 $K$, and radius = 26 $R_\odot$. For both stars we adopt the same all-electron spectrum that fits AMS-02 CR data at 1AU distance from the star with  $\Phi_{0}$ = 600MV modulation potential at 1~AU distance from the star. $\Phi(r)$ in any $r$ distance from the star is given by eq.~\ref{eq3}, with $\Phi_{r}$ = 0 MV for $r \geq r_b$. This means that we assume the local interstellar spectrum to be representative of the spectrum in the interstellar space, while assuming a pretty strong CR modulation in the vicinity of the star.   
The left plot in Figure~\ref{fig9} shows the cumulative flux from $\kappa$~Ori (brown lines) and $\zeta$~Pup (orange lines) as a function of integration angle for energies above 100~MeV (solid lines) and 1~GeV (dashed lines). 
The right plot in the same figure shows the integrated intensity above 100~MeV from $\kappa$~Ori and $\zeta$~Pup as a function of angular distance from the stars. The IGRB \cite{IGRB}, which represents all the isotropic emission currently detected by Fermi-LAT, is shown for comparison.

Figure~\ref{fig10} reports the calculated IC spectral flux of $\kappa$~Ori and $\zeta$~Pup in the energy range of Fermi-LAT for the AMS-02 spectrum with $\Phi_{0}$ = 600 MV in eq.~\ref{eq3} and integrated over areas with various angular amplitudes: 1$^\circ$, 4$^\circ$, and 8$^\circ$. As for the case of the Sun, spectral fluxes correspond to extension table 8 in the output FITS file (see Table~\ref{Table1}). These models are based on the CR electron flux as measured at Earth's proximity, which may not be the same in different locations in the Galaxy. Hence the gamma-ray flux from these stars can be even larger, increasing linearly with the density of the CR electrons. Detecting gamma-ray emission from stars other than our Sun can inform on CR electrons close to the location of the various stars, and even a non detection can put constraints on the CR electron spectrum in different locations in the Galaxy. The 10 years Fermi-LAT point source sensitivity is also shown\footnote{https://www.slac.stanford.edu/exp/glast/groups/canda/lat\_Performance.htm}. Following the official Fermi-LAT performance data, this sensitivity is given for (l,~b) = (120$^\circ$,~45$^\circ$) and for P8R3\_V2. (The reader should be aware that the Fermi-LAT sensitivity was obtained for a point source, while the IC component is an extended source).

\section{Conclusions}
Models for the solar IC emission are important for assisting analyses and for interpreting recent and forthcoming observations of the Sun from keV to TeV energies for various solar conditions. 
IC models from stars are also used for interpreting data at GeV energies. \\
In this work, we have presented the StellarICS code for calculating the IC emission from the Sun and stars. It provides computations accounting for various electron and positron spectral models, the isotropic or anisotropic Klein-Nishina formulation, various treatments for the solar (stellar) modulation, given technical parameters, such as the energy grid and angular steps, and given physical ones, such as distances, radius, temperatures of the Sun (stars), that can be freely chosen by the user. The code is publicly available and can be easily extended by the user to include additional and more sophisticated models. \\
After describing the code, we have presented examples of updated solar IC models for the Fermi-LAT energy band based on the various CR electron measurements. In particular, we have shown expectations for two baseline models, one for solar minimum conditions mainly based on Pamela CR measurements, and one for solar maximum conditions mainly based on AMS-02 CR measurements.
Fermi-LAT observes the Sun with high statistical significance. Current observations include data for the entire Cycle 24 for different solar conditions and polarity. Comparisons of data with models allow us to gain information on the leptonic and hadronic CRs and their modulation throughout the heliosphere and close to the Sun, and they allow studies of the solar environment, including transport and interaction even at the solar surface and over the different solar cycles with changes in polarity. \\
Besides updating the IC solar models in the Fermi-LAT energy range, we have also extended our calculations down to keV energies. %The proposed MeV mission such as AMEGO, GECCO, and e-ASTROGAM will achieve a major gain in sensitivity compared to the COMPTEL and INTEGRAL missions. 
We have estimated the solar extended IC emission to be very important for future MeV missions (e.g. AMEGO, e-Astrogam-like, and possibly GECCO), as it is for Fermi-LAT. Indeed, for the IC below 100~MeV the solar modulation effect is at its maximum, thus allowing to easily distinguish among different models. This will allow access for the first time the low-energy CR all-electrons close to the Sun. Furthermore, the solar IC emission will contribute to the observed MeV diffuse emission even at large angular distances from the Sun, as for Fermi-LAT. On the contrary, we have shown that below 100~keV the substantially higher contribution from the extragalactic x-ray background would make  the solar large-scale emission hardly detectable even by present sensitive x-ray survey telescopes, such as eRosita, being the expected IC solar emission many orders of magnitude below the extragalactic background. \\
Additionally, we have extended our calculations of the solar IC emission to TeV energies. Indeed the Sun  has recently started to be investigated also at these energies. TeV gamma-ray observations from the Sun (e.g.  from HAWC, LHAASO, and SWGO) may provide access to the all-electron spectrum above 20~TeV, which at the moment is not accessible to any present instruments.
We have accounted for an extreme CR all-electron spectrum to provide an upper limit to the expected IC emission. We found that current upper limits from the Sun with HAWC are consistent  with these IC calculations. 
By comparing this extreme model with the expected sensitivity of HAWC, LHAASO, and SWGO in the next years, we found that the solar IC emission could be barely detectable by these telescopes. This would require the flux of the all-electrons $\sim$20~TeV to be comparable to the flux of protons, which would be totally unexpected. \\
Finally, we have given calculations from a sample of closest luminous stars. 
Single stars and associations produce gamma rays by the same IC mechanism. 
This IC emission is expected to be detected by Fermi-LAT from the closest most luminous stars and for OB associations. We expect the smooth large-scale interstellar IC emission to be clumpy in these directions. Comparison of gamma-ray observations in the direction of stars with stellar models produced with the StellarICS code will also help to probe the all-electron spectrum in the proximity of the stars and thus at different positions in the Galaxy.

%\appendix
%\section{None}

\acknowledgments
E.O. acknowledges the ASI-INAF agreement n. 2017-14-H.0 
and the NASA Grant \\
No. 80NSSC20K1558. The authors thank AMEGO, e-ASTROGAM, HAWC, SWGO, and LHAASO teams for providing the data points of the sensitivities reported in the plots. The authors acknowledge the anonymous referee for constructive comments.
%This is the most common positions for acknowledgments. A macro is
%available to maintain the same layout and spelling of the heading.

%\paragraph{Note added.} This is also a good position for notes added
%after the paper has been written.

% The bibliography will probably be heavily edited during typesetting.
% We'll parse it and, using the arxiv number or the journal data, will
% query inspire, trying to verify the data (this will probalby spot
% eventual typos) and retrive the document DOI and eventual errata.
% We however suggest to always provide author, title and journal data:
% in short all the informations that clearly identify a document.

\newcommand{\mnras} {MNRAS}
\newcommand{\apjl}{ApJL}
\newcommand{\apj}{ApJ}
\newcommand{\ssr}{Space Sci. Rev.}
\newcommand{\apjs}{ApJS}
\newcommand{\prd}{PhRvD}
\newcommand{\aap}{A\&A}
\newcommand{\nat}{Nature}
\newcommand{\jcap}{JCAP}
\newcommand{\aaps}{A\&AS}
\newcommand{\prl}{PRL}
\newcommand{\baas}{Bulletin of the American Astronomical Society}
\newcommand{\jgr}{Journal of Geophysical Research}

\end{document}